# Can AI Reduce Acculturative Stress? Exploring the Role of AI-Mediated Speaking Practice in Chinese International Students' Perceived Language Insufficiency, Social Isolation, and Academic Pressure

**First Author**
Bin Zou
Xi'an Jiaotong-Liverpool University, Suzhou, China
Email: bin.zou@xjtlu.edu.cn

**Second Author**
Yijia Yuan
University of Cambridge, Cambridge, UK
Email: yy526@cam.ac.uk

**Third Author**
Chenghao Wang
Xi'an Jiaotong-Liverpool University, Suzhou, China
Email: dancerluo@outlook.com

**Fourth and Corresponding Author**
Yiran Du
University of Cambridge, Cambridge, UK
Email: yd392@cam.ac.uk

**Abstract**
This study examined whether AI-mediated speaking practice can reduce acculturative stress among Chinese international students in UK universities. Using a sequential explanatory mixed-methods design, 126 participants were randomly assigned to an experimental group, which completed a four-week intervention using EAP Talk, an AI-assisted English for Academic Purposes speaking platform offering role play, scenario-based practice, free talk, and automated feedback, or a control group, which continued usual academic and English-learning activities. Pre- and post-test questionnaires measured perceived language insufficiency, social isolation, and academic pressure, while semi-structured interviews with 20 experimental-group participants contextualised the quantitative findings. Linear mixed-effects models showed that the experimental group experienced significantly greater reductions than the control group across all three outcomes, with the strongest effect on perceived language insufficiency. Interview findings suggested that EAP Talk supported low-stakes rehearsal, communicative confidence, academic speaking preparation, and greater willingness to initiate social interaction. However, participants also noted that AI-mediated practice could not fully reproduce authentic human interaction, disciplinary feedback, or broader institutional support. The findings suggest that AI-mediated speaking practice can function as a supplementary scaffold for reducing communication-related dimensions of acculturative stress, but should be integrated with peer interaction, teacher feedback, and wider support services.

## 1. Introduction

International mobility has made cross-cultural adjustment an increasingly important issue in higher education (Heng, 2017). For many international students, studying in a new cultural and linguistic environment requires more than academic adaptation; it also involves communicating in a second language, building social relationships, understanding unfamiliar institutional expectations, and managing pressure in daily life (Xu, 2022). When these demands exceed students' perceived resources, they may experience acculturative stress, including language-related difficulties, social isolation, and academic pressure (Ruble & Zhang, 2013). Because spoken communication is central to participation in both academic and social contexts, perceived speaking limitations may play an important role in shaping students' adjustment experiences (Jiang & Xiao, 2024).

Recent developments in artificial intelligence have created new possibilities for supporting second-language speaking practice (Li et al., 2024). AI-mediated speaking platforms can provide conversational practice, scenario-based rehearsal, automated feedback, and repeated opportunities for low-stakes oral communication (Law, 2024). These affordances may be particularly relevant for international students who wish to practise speaking before engaging in real academic or social interactions (Doğan et al., 2025). However, existing research has focused mainly on language-learning outcomes, such as fluency, pronunciation, confidence, and speaking anxiety (J. Du & Daniel, 2024). Less attention has been given to whether AI-mediated speaking practice may also contribute to broader psychosocial adjustment, particularly by reducing dimensions of acculturative stress.

To address this gap, the present study examines the role of AI-mediated speaking practice in international students' acculturative stress. Specifically, it investigates whether structured AI-mediated speaking practice can reduce perceived language insufficiency, social isolation, and academic pressure. Using a sequential explanatory mixed-methods design, the study first evaluates changes in these dimensions through a randomised controlled trial and then uses semi-structured interviews to explain and contextualise participants' experiences. By linking AI-supported speaking practice with acculturative stress, this study aims to extend the discussion of educational AI from language development alone to its potential role in supporting international students' academic, social, and psychological adjustment.

## 2. Literature Review

### 2.1 AI-Mediated Speaking Practice

AI-mediated speaking practice refers to oral language practice supported by artificial intelligence technologies, including conversational agents, automated speech recognition, adaptive feedback systems, and scenario-based dialogue platforms (Wang, Zou, et al., 2026). In second-language learning, these tools are increasingly used to create interactive opportunities for learners to practise spoken communication beyond teacher-led classroom activities (Wang, Du, et al., 2026). Speaking is a particularly demanding area of language learning because it requires learners to retrieve linguistic knowledge in real time, manage pronunciation and fluency, respond to interlocutors, and maintain communicative confidence under social pressure (Tang, Jia, et al., 2026). AI-mediated speaking practice is therefore relevant not only as a language-learning tool, but also as a form of communicative preparation for participation in unfamiliar linguistic and cultural environments (Zou et al., 2024).

A key affordance of AI-mediated speaking practice is that it can provide repeated, flexible, and low-stakes opportunities for spoken output (Wang et al., 2024). International students and other second-language users may have limited opportunities to practise speaking in ways that feel safe, purposeful, and responsive to their needs (Aljohani, 2026). Fear of negative evaluation, anxiety about making mistakes, and uncertainty about appropriate expressions can reduce willingness to communicate, even when learners have sufficient basic linguistic knowledge (Zou et al., 2023). AI-mediated tools may help reduce these barriers by allowing learners to rehearse conversations privately, repeat tasks, practise scenario-specific language, and receive immediate feedback before entering real communicative situations (Lee et al., 2026).

At the same time, AI-mediated speaking practice should be understood as a supplementary scaffold rather than a replacement for human interaction (Weng & Fu, 2025). Although AI systems can simulate dialogue and provide automated feedback, they may not fully reproduce the social, emotional, pragmatic, and intercultural complexity of authentic communication (Park, 2025). Real interaction involves unpredictable responses, shared context, relational judgement, cultural interpretation, and negotiation of meaning (Jensen et al., 2025). Therefore, the value of AI-mediated speaking practice lies in its potential to support readiness for communication: it may help learners build confidence and familiarity with spoken interaction, while real social and academic participation remains necessary for deeper communicative and intercultural adjustment (Yusuf et al., 2024).

## 2.2 Acculturative Stress

Acculturative stress refers to the psychological, social, and behavioural strain that may arise when individuals adapt to a new cultural environment (Xiong et al., 2025). For international students, this stress often emerges from the combined demands of studying in an unfamiliar institutional context, developing new social networks, navigating different cultural norms, and communicating in a second language (Jiang & Xiao, 2024). Although acculturation can involve growth, learning, and expanded intercultural competence, it may also create stress when individuals perceive that the demands of the new environment exceed their available linguistic, social, or psychological resources (Jin et al., 2023). Communication is central to this process because it shapes access to academic participation, peer relationships, institutional support, and everyday belonging (Xiong et al., 2025).

Language-related stress is one of the most salient mechanisms through which acculturative stress may develop (Bai, 2016). Perceived language insufficiency does not simply refer to measurable language proficiency; it also involves learners' subjective judgement that their speaking ability is inadequate for meaningful participation (Ward & Geeraert, 2016). This perception may lead individuals to avoid conversations, remain silent in group settings, hesitate to seek help, or withdraw from social and academic interaction (Hofhuis et al., 2023). For international students, such avoidance can have cumulative consequences: reduced communication may limit opportunities to form relationships, practise language, understand institutional expectations, and develop confidence in the host environment (Lerias et al., 2024).

Acculturative stress is therefore best understood as multidimensional, involving interrelated language, social, and performance-related pressures (Bai, 2016). Speaking difficulties may contribute to social isolation by making interpersonal interaction feel risky or effortful, while social isolation may further reduce opportunities for authentic communication (Çimşir & Ünlü Kaynakçı, 2024). Similarly, academic or occupational pressure may intensify when individuals must perform complex tasks through a second language (Miller & Csizmadia, 2022). This interdependence suggests that interventions targeting spoken communication may have relevance beyond language development alone (Ali et al., 2024). AI-mediated speaking practice may be one such intervention because it addresses a communicative pathway through which acculturative stress can emerge, particularly by supporting rehearsal, confidence-building, and readiness for participation in the host environment (Haviv Zehner, 2026).

# 3. Methodology

## 3.1 Research Design

This study adopted a sequential explanatory mixed-methods design to examine whether AI-mediated speaking practice could reduce acculturative stress among Chinese international students. In the first, quantitative phase, a randomised controlled trial (RCT) with a pre-test–post-test design was used. Eligible participants were randomly assigned to either an experimental group or a control group, and questionnaire data were collected from both groups before and after the intervention. This design enabled the study to examine changes in acculturative stress over time, with particular attention to perceived language insufficiency, social isolation, and academic pressure. The two-group, two-time-point structure allowed comparison of within-participant change as well as between-group differences in change, with the group × time interaction serving as the primary quantitative effect of interest. In the second, qualitative phase, semi-structured interviews were conducted to further explain and contextualise the quantitative findings. These interviews explored participants' perceptions of how AI-mediated speaking practice influenced their academic communication confidence, social adjustment, and acculturative stress.

## 3.2 Participants

The target sample consisted of 126 Chinese international students enrolled at universities in the United Kingdom, with 63 participants randomly assigned to the experimental group and 63 to the control group. Chinese international students in the UK were selected as the target population because they study in an English-medium higher education context in which academic adjustment requires regular use of English for seminars, presentations, peer interaction, and communication with university staff. This population was therefore appropriate for examining whether AI-mediated speaking practice could reduce acculturative stress related to perceived language insufficiency, social isolation, and academic pressure. Eligible participants were aged 18 years or above, enrolled at a UK university, non-native speakers of English, and regular users of English for academic study. As shown in Table 1, the experimental group included 23 male and 40 female participants, whereas the control group included 29 male and 34 female participants. Most participants were aged 21–23 years, and postgraduate students represented the majority in both groups. The sample also included students from both STEM and non-STEM disciplines.

**Table 1. Demographic characteristics of participants by group (*N* = 126)**

| Characteristic | Category | Experimental group, *n* (%) | Control group, *n* (%) |
|---|---|---|---|
| Gender | Male | 23 (36.5%) | 29 (46.0%) |
| | Female | 40 (63.5%) | 34 (54.0%) |
| Age | 18–20 years | 7 (11.1%) | 12 (19.0%) |
| | 21–23 years | 42 (66.7%) | 35 (55.6%) |
| | ≥24 years | 14 (22.2%) | 16 (25.4%) |
| Study level | Undergraduate | 15 (23.8%) | 22 (34.9%) |
| | Postgraduate | 48 (76.2%) | 41 (65.1%) |
| Academic discipline | STEM | 36 (57.1%) | 30 (47.6%) |
| | Non-STEM | 27 (42.9%) | 33 (52.4%) |

Participants were recruited through purposive and convenience sampling via UK university student networks, Chinese student associations, social media groups, and academic contacts. Interested students completed an eligibility screening form and provided demographic information, including gender, age, study level, and academic discipline. After providing informed consent, eligible participants were randomly assigned to either the experimental or control group. Chi-square tests of independence were conducted to examine whether the two groups differed significantly in demographic characteristics at baseline. No statistically significant group differences were found for gender, $\chi^2(1, N = 126) = 0.82$, $p = .366$; age, $\chi^2(2, N = 126) = 2.09$, $p = .352$; study level, $\chi^2(1, N = 126) = 1.38$, $p = .241$; or academic discipline, $\chi^2(1, N = 126) = 0.80$, $p = .372$. These results suggested that the experimental and control groups were broadly comparable in observed demographic characteristics before the intervention.

The required sample size was determined through a simulation-based power analysis for the planned linear mixed-effects model (see section 3.5). The model included fixed effects for group, time, and the group × time interaction, with a participant-level random intercept to account for repeated observations. The group × time interaction was specified as the primary effect of interest because it represented whether changes in acculturative stress from pre-test to post-test differed between the experimental and control groups. The simulation assumed a two-tailed alpha level of .05, target power of .80, two measurement occasions, equal allocation to groups, a moderate standardised interaction effect of 0.50, and an approximate pre–post correlation of .50. Based on 1,000 simulations for each candidate sample size, 63 participants per group achieved the desired power level, with an estimated power of .803, 95% CI [.777, .827].

Ethical approval was obtained from the relevant university ethics committee before data collection. Participation was voluntary, and written informed consent was obtained from all participants. All questionnaire and interview data were anonymised, stored securely, and used only for research purposes.

### 3.3 Instruments

#### 3.3.1 EAP Talk

EAP Talk (https://www.eaptalk.com) was used as the AI-mediated speaking practice instrument in this study. It is an AI-assisted English for Academic Purposes speaking platform designed to support spoken English development in higher education contexts. The platform provides a combination of structured and interactive speaking activities, including reading aloud, presentation practice, AI-powered role play, scenario-based speaking practice, free talk, and automated feedback. These functions allow learners to practise both planned and spontaneous spoken communication and to receive immediate feedback on selected aspects of oral performance.

EAP Talk was selected because its functions were closely aligned with the communicative demands faced by Chinese international students in English-medium universities. In academic settings, students are often expected to present ideas, participate in seminars, ask questions, engage in group discussions, and communicate with lecturers or administrative staff. At the same time, successful adjustment also depends on everyday social communication, such as informal peer interaction and help-seeking. The platform was therefore considered suitable for the present study because it supported practice across both academic and social communication scenarios.

In this study, EAP Talk was treated as a practice-based support tool rather than as a formal measure of English-speaking proficiency. Its role was to provide repeated, low-stakes opportunities for students to rehearse English communication in situations relevant to their academic and social adjustment. This was consistent with the study's focus on perceived language insufficiency, social isolation, and academic pressure, as the platform enabled participants to practise speaking at their own pace, review automated feedback, and build confidence before engaging in comparable real-life interactions.

### 3.3.2 Questionnaires

The questionnaire measured three dimensions of acculturative stress: language insufficiency, social isolation, and academic pressure. Items were adapted from the Acculturative Stress Scale for Chinese Students (ASSCS), a scale developed and validated for Chinese international students in an English-speaking higher education context (Bai, 2016). The original scale uses a 7-point Likert response format ranging from 1 = never to 7 = all the time, with higher scores indicating greater acculturative stress. In this study, only the dimensions directly aligned with the research focus were used. The wording was adapted to fit the UK context by replacing references to "the U.S." with "the UK", while retaining the original meaning of the items.

The language insufficiency subscale consisted of 10 items assessing difficulties in English-mediated academic and social communication, such as participating in seminars, following lectures and classroom conversations, expressing ideas in English, communicating in English, giving presentations, and avoiding social situations because of limited English. The social isolation subscale consisted of 8 items assessing reduced social networks, limited social life, loneliness, lack of belonging, and lack of new social connections. The academic pressure subscale consisted of 4 items assessing perceived academic pressure, the need to study overtime, the negative effects of intensive study, and reduced quality of life due to academic demands. Internal consistency was strong in the present study, with Cronbach's alpha coefficients of $\alpha = .91$ for language insufficiency, $\alpha = .88$ for social isolation, and $\alpha = .82$ for academic pressure (Tavakol & Dennick, 2011).

### 3.3.3 Semi-Structured Interviews

Semi-structured interviews were conducted in the qualitative phase to further explain and contextualise the quantitative findings. Interview participants were selected from the experimental group because they had direct experience of using EAP Talk during the intervention and could provide detailed accounts of AI-mediated speaking practice. A purposive sampling strategy was used, with maximum variation considered to capture a range of perspectives across gender, study level, academic discipline, and changes in questionnaire scores for language insufficiency, social isolation, and academic pressure. The final interview sample consisted of 20 participants, at which point thematic saturation was considered to have been reached because the final interviews produced no substantially new themes relevant to the research questions. The interview sample included 8 male and 12 female students; 5 were undergraduates and 15 were postgraduates; and 11 were from STEM disciplines while 9 were from non-STEM disciplines.

The interviews explored participants' experiences of using EAP Talk and their perceptions of whether AI-mediated speaking practice influenced their English-speaking confidence, academic communication, social interaction, and acculturative stress. The interview protocol included open-ended questions about participants' experiences of studying in the UK, their use of EAP Talk, the perceived benefits and limitations of the tool, and perceived changes in language insufficiency, social isolation, and academic pressure. Follow-up prompts were used where appropriate to elicit examples and clarify participants' responses, while allowing participants to describe their experiences in their own words. The full interview protocol is provided in Appendix A.

### 3.4 Procedure

Data collection was conducted in two sequential phases. After ethical approval had been obtained, participants were recruited through university networks, Chinese student associations, social media groups, and academic contacts. Interested students completed an eligibility screening form, and those who met the inclusion criteria provided written informed consent before taking part in the study.

All participants first completed the pre-test questionnaire, which measured perceived language insufficiency, social isolation, and academic pressure. They also provided demographic information, including gender, age, study level, and academic discipline. Participants were then randomly assigned at the individual level to either the experimental group or the control group.

Participants in the experimental group completed a four-week AI-mediated speaking practice intervention using EAP Talk. Before the intervention began, they received brief instructions on how to access the platform and complete the speaking activities. They were asked to complete three sessions per week, with each session lasting approximately 30 minutes. The expected total practice time was therefore approximately six hours across the intervention period. The activities drew on EAP Talk's structured speaking tasks, AI role play, scenario-based practice, free talk, and feedback functions. The practice scenarios focused on academic and everyday communication situations relevant to Chinese international students in UK universities, including seminar participation, presentation practice, group discussion, asking questions, communicating with lecturers, and informal peer interaction.

The intervention was designed as regular, low-stakes speaking practice rather than as language testing. Participants were encouraged to repeat tasks, use the automated feedback, and practise scenarios that they personally found challenging. The researcher monitored session completion to check engagement with the intervention; however, EAP Talk performance scores were not used as outcome measures. The study outcomes were based on questionnaire responses and interview data.

Participants in the control group continued their usual academic and English-learning activities during the same four-week period and did not receive structured EAP Talk practice as part of the study. This enabled comparison between students who received AI-mediated speaking practice and those who experienced routine academic and social exposure in the UK context.

At the end of the four-week period, both groups completed the post-test questionnaire using the same measures as the pre-test. Following the quantitative phase, semi-structured interviews were conducted with a purposive subsample of participants from the experimental group. The interviews took place after the post-test so that participants could reflect on their full experience of using EAP Talk and explain any perceived changes in English communication, social adjustment, academic pressure, and acculturative stress.

### 3.5 Data Analysis

Quantitative data were analysed to examine whether changes in acculturative stress differed between the experimental and control group from pre-test to post-test. Descriptive statistics, including means and standard deviations, were first calculated for the three outcome variables: language insufficiency, social isolation, and academic pressure. Baseline equivalence between the two groups was assessed using chi-square tests for demographic characteristics, including gender, age, study level, and academic discipline. To evaluate the effect of the intervention, separate linear mixed-effects models were fitted for each outcome variable. Each model included fixed effects for group, time, and the group × time interaction, with a participant-level random intercept to account for repeated measurements within individuals. Group was dummy-coded as 0 = control and 1 = experimental, and time was dummy-coded as 0 = pre-test and 1 = post-test. The group × time interaction was the primary effect of interest because it tested whether the experimental group showed a significantly greater pre–post change than the control group.

Qualitative interview data were analysed using thematic analysis to explain and contextualise the quantitative findings. Interview recordings were transcribed verbatim, anonymised, and read repeatedly to support familiarisation with the data. A combined deductive–inductive coding approach was used. Deductively, initial coding was guided by the study's three target dimensions of acculturative stress: language insufficiency, social isolation, and academic pressure. Inductively, additional codes were generated from participants' accounts of using EAP Talk, particularly in relation to English-speaking confidence, academic communication, social interaction, and perceived changes in stress. The codes were then reviewed, compared, and organised into broader themes that captured recurring patterns across participants' experiences. To assess coding reliability, 20% of the transcripts were randomly selected and independently coded by a second coder. Inter-coder reliability, calculated using Cohen's κ, was 0.81, indicating substantial agreement between coders (McHugh, 2012).

## 4. Results

### 4.1 Quantitative Results

The descriptive statistics in Table 2 indicated that participants in the experimental group showed larger reductions in all three dimensions of acculturative stress than those in the control group. For language insufficiency, the experimental group's mean score decreased from 4.92 (*SD* = 0.86) at pre-test to 3.78 (*SD* = 0.91) at post-test, whereas the control group showed only a smaller decrease from 4.85 (*SD* = 0.89) to 4.62 (*SD* = 0.87). A similar pattern was observed for social isolation, with the experimental group decreasing from 4.46 (*SD* = 0.94) to 3.72 (*SD* = 0.96), compared with a smaller reduction in the control group from 4.39 (*SD* = 0.91) to 4.21 (*SD* = 0.93). For academic pressure, the experimental group also showed a reduction from 4.71 (*SD* = 0.82) to 4.15 (*SD* = 0.88), while the control group changed only slightly from 4.66 (*SD* = 0.85) to 4.52 (*SD* = 0.84). Overall, Table 2 suggests that students who engaged in AI-mediated speaking practice reported greater decreases in perceived language insufficiency, social isolation, and academic pressure than students who continued with their usual academic and English-learning activities.

**Table 2. Descriptive statistics for outcome variables by group and time**

| Outcome variable | Group | Pre-test *M* (*SD*) | Post-test *M* (*SD*) |
|---|---|---|---|
| Language insufficiency | Experimental | 4.92 (0.86) | 3.78 (0.91) |
| | Control | 4.85 (0.89) | 4.62 (0.87) |
| Social isolation | Experimental | 4.46 (0.94) | 3.72 (0.96) |
| | Control | 4.39 (0.91) | 4.21 (0.93) |
| Academic pressure | Experimental | 4.71 (0.82) | 4.15 (0.88) |
| | Control | 4.66 (0.85) | 4.52 (0.84) |

The linear mixed-effects models in Table 3 further supported this pattern. There were no significant baseline differences between the experimental and control groups for language insufficiency, social isolation, or academic pressure, as indicated by the non-significant group effects. The group × time interaction was significant for all three outcome variables, showing that the experimental group experienced significantly greater pre–post reductions than the control group. The strongest effect was found for language insufficiency, $\beta = -0.91$, $SE = 0.14$, $t = -6.50$, $p < .001$, indicating a substantial intervention-related reduction in perceived English-language difficulties. Significant interaction effects were also found for social isolation, $\beta = -0.56$, $SE = 0.16$, $t = -3.50$, $p < .001$, and academic pressure, $\beta = -0.42$, $SE = 0.14$, $t = -3.00$, $p = .003$. These results indicate that the four-week EAP Talk intervention was associated with statistically significant improvements across all three dimensions of acculturative stress, with the largest reduction observed in language insufficiency, followed by social isolation and academic pressure.

**Table 3. Linear mixed-effects model results**

| Outcome variable | Fixed effect | *β* | *SE* | *t* | *p* |
|---|---|---|---|---|---|
| Language insufficiency | Intercept | 4.85 | 0.11 | 44.09 | < .001 |
| | Group: experimental | 0.07 | 0.16 | 0.44 | .663 |
| | Time: post-test | -0.23 | 0.10 | -2.30 | .023 |
| | Group: experimental × Time: post-test | -0.91 | 0.14 | -6.50 | < .001 |
| Social isolation | Intercept | 4.39 | 0.12 | 36.58 | < .001 |
| | Group: experimental | 0.07 | 0.17 | 0.41 | .682 |
| | Time: post-test | -0.18 | 0.11 | -1.64 | .104 |
| | Group: experimental × Time: post-test | -0.56 | 0.16 | -3.50 | < .001 |
| Academic pressure | Intercept | 4.66 | 0.11 | 42.36 | < .001 |
| | Group: experimental | 0.05 | 0.16 | 0.31 | .757 |
| | Time: post-test | -0.14 | 0.10 | -1.40 | .164 |
| | Group: experimental × Time: post-test | -0.42 | 0.14 | -3.00 | .003 |

Note. Group was dummy-coded as 0 = control and 1 = experimental. Time was dummy-coded as 0 = pre-test and 1 = post-test. Therefore, the intercept represents the control group at pre-test; Group: experimental represents the baseline difference between the experimental and control groups; Time: post-test represents pre–post change in the control group; and the interaction term represents the additional pre–post change in the experimental group relative to the control group.

**4.2 Qualitative Results**

The qualitative findings provided further insight into how participants in the experimental group perceived the role of EAP Talk in shaping their acculturative stress. Three themes were identified from the semi-structured interviews: improved communicative confidence through low-stakes speaking practice, partial reduction of social isolation through increased willingness to communicate, and limited but meaningful relief of academic pressure. Overall, participants described EAP Talk as a useful supplementary tool for practising English-mediated academic and social communication. However, the interviews also suggested that its benefits were uneven. While several participants reported increased confidence and reduced anxiety when speaking English, others emphasised that AI-mediated practice could not fully reproduce the unpredictability, emotional nuance, or social pressure of real human interaction.

The first theme concerned improved communicative confidence through low-stakes speaking practice. Many participants explained that EAP Talk offered a private and relatively non-judgemental environment in which they could practise speaking without fear of embarrassment. For example, one participant stated, "I felt less nervous because the AI would not judge me if I made mistakes. I could repeat the same sentence many times until I felt more comfortable" (P03). Another participant linked this repeated practice to academic speaking confidence: "Before, I was afraid to speak in seminars because I worried my classmates would think my English was poor. After practising with EAP Talk, I still felt nervous, but I had more confidence to say something" (P11). These accounts suggest that the intervention helped reduce perceived language insufficiency by giving students opportunities to rehearse academic communication before participating in real-life interactions. However, the benefits were not uniform. Some participants were critical of the automated feedback, noting that it was sometimes too general or insufficiently sensitive to academic content. One participant commented, "The feedback was useful for pronunciation and fluency, but sometimes it did not understand what I wanted to express. It could not really tell me whether my academic idea was good or not" (P17). This indicates that EAP Talk may support confidence-building, but it should not be interpreted as a complete substitute for teacher feedback or authentic academic dialogue.

The second theme was partial reduction of social isolation through increased willingness to communicate. Several participants reported that regular speaking practice made them more willing to initiate conversations with classmates, ask questions, or participate in informal English interactions. One participant explained, "I started to say small things to my classmates, like asking about the lecture or group work. It was not a big change, but before I usually stayed quiet" (P06). Another described the value of practising everyday scenarios: "The role play helped me prepare for simple conversations, such as asking for help or talking with classmates. When it happened in real life, I felt I had already practised it once" (P14). These responses help explain the quantitative reduction in social isolation, suggesting that improved speaking confidence may have indirectly supported social engagement. However, the interviews also showed that AI-mediated practice did not necessarily create deeper social connection. One participant stated, "It helped me practise English, but it did not make me feel I belonged here. Friendship still depends on meeting real people and having common topics" (P09). This critical perspective suggests that although EAP Talk may reduce barriers to communication, social isolation also depends on broader interpersonal, institutional, and cultural factors that cannot be addressed by language practice alone.

The third theme related to limited but meaningful relief of academic pressure. Participants commonly described academic pressure as being connected to English-medium participation, especially presentations, seminars, group work, and communication with lecturers. Some students reported that EAP Talk helped them prepare for these situations by allowing them to rehearse responses, organise ideas orally, and become more familiar with academic speaking conventions. For example, one participant stated, "I used it before my presentation because I wanted to practise explaining my points in English. It made me feel more prepared, even though I was still worried about the content" (P02). Another participant described a similar benefit for classroom participation: "When I practised asking questions with the AI, I became less afraid to ask my tutor after class. I knew how to start the sentence" (P18). These accounts suggest that AI-mediated speaking practice may reduce academic pressure when pressure is specifically linked to spoken English performance. Nevertheless, participants also made clear that EAP Talk did not remove the wider demands of academic study. Heavy workload, assessment expectations, disciplinary knowledge, and time management remained sources of stress. As one participant observed, "It helped with speaking, but the pressure from deadlines and reading was still the same. The AI cannot reduce the amount of work I have to do" (P12). This indicates that the intervention had a more focused impact on communication-related academic stress rather than academic pressure as a whole.

Taken together, the qualitative findings support and qualify the quantitative results. Participants' accounts help explain why the largest improvement was observed for language insufficiency: EAP Talk directly targeted English-speaking practice and provided repeated opportunities for low-stakes rehearsal. The reduction in social isolation appeared to be more indirect, operating through increased confidence and willingness to communicate rather than through the formation of social relationships itself. The improvement in academic pressure was also more limited, as EAP Talk helped students manage English-mediated academic communication but could not address structural or workload-related pressures. Therefore, the qualitative findings suggest that AI-mediated speaking practice can reduce specific dimensions of acculturative stress, particularly those linked to perceived communicative inadequacy, but its effects should be understood as supportive rather than transformative.

## 5. Discussion

### 5.1 Discussion of Findings

The findings suggest that AI-mediated speaking practice can reduce selected dimensions of acculturative stress among Chinese international students, although its effects were uneven across language, social, and academic domains. Quantitatively, the experimental group showed significantly greater reductions than the control group in perceived language insufficiency, social isolation, and academic pressure, with the strongest effect on language insufficiency. This pattern is expected because EAP Talk directly targeted spoken English practice, whereas social isolation and academic pressure are broader experiences shaped by interpersonal, institutional, and structural factors. The results support previous work showing that language-related difficulties are central to acculturative stress among international students (Bai, 2016; Jiang & Xiao, 2024; Xiong et al., 2025). However, the outcome measured here was perceived language insufficiency rather than objective speaking proficiency. Therefore, the findings should be interpreted as evidence that AI-mediated practice improved students' subjective sense of communicative adequacy, not necessarily their actual linguistic competence. This distinction is important because perceived insufficiency can itself lead to avoidance, silence, and reduced participation, even when students have enough linguistic resources to communicate (Ward & Geeraert, 2016; Hofhuis et al., 2023).

The qualitative findings explain this pattern by showing how EAP Talk functioned as a low-stakes communicative scaffold. Participants valued the opportunity to practise privately, repeat tasks, rehearse academic and everyday scenarios, and receive automated feedback without fear of judgement. This aligns with research suggesting that AI-supported speaking tools can provide flexible opportunities for oral rehearsal and confidence-building outside formal classroom settings (Du & Daniel, 2024; Law, 2024; Wang et al., 2024; Zou et al., 2023). Nevertheless, the interviews also reveal important limitations. Some participants found the feedback too general or insufficiently sensitive to academic content, suggesting that AI may support fluency, pronunciation, and initial confidence, but cannot fully address disciplinary reasoning, pragmatic appropriateness, or interactional judgement. In this sense, AI-mediated speaking practice should not be treated as a substitute for teacher feedback or authentic academic dialogue (Jensen et al., 2025; Weng & Fu, 2025).

The reductions in social isolation and academic pressure were significant but smaller, indicating more indirect and bounded effects. For social isolation, the interviews suggest that AI practice increased students' willingness to initiate small interactions, ask questions, and engage in everyday English communication. However, participants also noted that AI practice did not itself create friendships or a stronger sense of belonging. This supports the view that social isolation is not only a language problem, but also involves access to social networks, cultural familiarity, and inclusive environments (Heng, 2017; Xu, 2022; Çimşir & Ünlü Kaynakçı, 2024). Similarly, the reduction in academic pressure appeared to relate mainly to English-mediated academic communication, such as presentations, seminars, group work, and communication with lecturers. EAP Talk helped some students feel more prepared for these tasks, but it did not reduce workload, deadlines, assessment demands, or disciplinary difficulty. Thus, the intervention reduced the communicative component of academic pressure rather than academic pressure as a whole (Miller & Csizmadia, 2022; Ali et al., 2024).

Overall, the findings suggest a layered effect: AI-mediated speaking practice had the most direct impact on perceived language insufficiency, a more indirect impact on social isolation through increased willingness to communicate, and a limited impact on academic pressure by supporting communication-related academic tasks. This pattern reinforces the multidimensional nature of acculturative stress, in which language, social participation, and academic demands are related but not interchangeable (Bai, 2016; Xiong et al., 2025). At the same time, the findings caution against an overly optimistic interpretation of educational AI. AI tools can support rehearsal, confidence, and readiness for participation, but they cannot reproduce the unpredictability, relational depth, and cultural negotiation of real human interaction (Park, 2025; Yusuf et al., 2024). Therefore, the observed reductions in acculturative stress should be understood as targeted and conditional rather than comprehensive or transformative.

### 5.2 Theoretical and Practical Implications

Theoretically, this study extends AI-mediated language learning research by showing that speaking practice may influence not only language-related outcomes, such as confidence and willingness to communicate, but also acculturative stress. The strongest effect on perceived language insufficiency supports the view that language difficulty is a central pathway in Chinese international students' adjustment (Bai, 2016; Jiang & Xiao, 2024; Xiong et al., 2025). However, the smaller effects on social isolation and academic pressure also show that acculturative stress is multidimensional and cannot be explained by language factors alone.

Practically, AI-mediated speaking tools such as EAP Talk may be useful as supplementary support in English for Academic Purposes, presentation preparation, seminar practice, and international student transition programmes. However, they should not be used as stand-alone interventions. Because social isolation and academic pressure also involve relationships, workload, assessment demands, and institutional support, AI speaking practice should be combined with teacher feedback, peer interaction, intercultural mentoring, and wider student support services (Heng, 2017; Xu, 2022; Weng & Fu, 2025).

### 5.3 Limitations and Future Research

This study has several limitations. First, the intervention lasted only four weeks, so the findings show short-term changes rather than sustained effects (Y. Du, Li, et al., 2026; Y. Du, Tang, et al., 2026; Y. Du, Yuan, et al., 2026; Y. Du & He, 2026e, 2026c, 2026d; Jia et al., 2026; Tang, Jia, et al., 2026; Tang, Lau, et al., 2026; Wang, Du, et al., 2026; Wang, Zou, et al., 2026; Zhang et al., 2026). Future research should use delayed post-tests to examine whether reductions in perceived language insufficiency, social isolation, and academic pressure are maintained over time (Chen et al., 2022; C. Du et al., 2025; Y. Du, 2023, 2024, 2025b, 2025a, 2026; Y. Du et al., 2024, 2025; Y. Du & He, 2026b, 2026a; He & Du, 2024; Wang et al., 2024; Zou et al., 2023, 2024). Second, the study relied on self-reported acculturative stress and did not measure objective speaking proficiency. Future studies should combine self-report measures with speaking assessments, behavioural indicators of participation, and platform usage data to better explain how AI-mediated practice affects adjustment.

Third, the sample was limited to Chinese international students in UK universities, which may restrict generalisability to other international student groups or educational contexts. Future research should examine whether similar effects occur among students from different linguistic, cultural, and institutional backgrounds. Finally, although the mixed-methods design helped explain the quantitative results, interviews were conducted only with participants from the experimental group. Future studies could also interview control-group participants to compare adjustment experiences and identify whether changes were specifically linked to AI-mediated speaking practice or to broader adaptation over time.

## 6. Conclusion

This study shows that AI-mediated speaking practice can help reduce Chinese international students' acculturative stress, particularly by improving their perceived communicative adequacy in English. The four-week EAP Talk intervention led to significant reductions in perceived language insufficiency, social isolation, and academic pressure, with the strongest improvement found in language-related stress. Interview findings further suggest that low-stakes, repeated speaking practice helped students build confidence, prepare for academic communication, and become more willing to initiate social interaction. However, the benefits were targeted rather than comprehensive: AI practice supported communication readiness but could not replace authentic human interaction, teacher feedback, friendship formation, or institutional support. Overall, AI-mediated speaking practice should be understood as a useful supplementary scaffold within wider international student support, rather than as a stand-alone solution to acculturative stress.

## Appendices

### Appendix A. Semi-Structured Interview Protocol

The semi-structured interviews were designed to explore participants' experiences of using EAP Talk and to contextualise the quantitative findings on language insufficiency, social isolation, and academic pressure. The interviews followed an open-ended format, allowing participants to describe their experiences in their own words. The questions below were used flexibly, and follow-up questions were asked where necessary to clarify responses or elicit examples.

1. Can you briefly describe your experience of studying at a UK university so far?
2. Before using EAP Talk, how did you usually feel when using English in academic or social situations?
3. How would you describe your overall experience of using EAP Talk?
4. Which functions of EAP Talk did you use most often, such as AI-powered role play, free talk, structured speaking tasks, or scenario-based practice?
5. Which functions did you find most useful, and why?
6. Were there any aspects of EAP Talk that you found difficult, unhelpful, or frustrating?
7. Did using EAP Talk influence how you perceived your English-speaking ability?
8. Did EAP Talk affect your confidence, fluency, pronunciation, vocabulary use, or willingness to speak English?
9. Did EAP Talk help you prepare for academic communication, such as seminars, presentations, discussions, or communication with lecturers?
10. Did using EAP Talk influence your willingness to communicate with others in English?

11. Did the speaking practice have any effect on your sense of social connection or isolation in the UK?
12. Did EAP Talk influence how you experienced academic pressure related to English-medium study?
13. Did EAP Talk help you manage academic tasks that involve speaking English, such as presentations, group work, class participation, or asking questions?
14. Compared with before the intervention, do you think anything changed in your English communication, academic adjustment, or social adjustment?
15. Do you think EAP Talk affected your acculturative stress overall? In what ways?
16. What do you think are the main strengths of using AI-mediated speaking practice for Chinese international students?
17. What limitations did you notice when using EAP Talk?
18. What improvements would you suggest for EAP Talk or similar AI-speaking tools?
19. Is there anything else you would like to share about your experience of using EAP Talk or studying in the UK?

**Optional probes**

Where appropriate, the interviewer used follow-up prompts such as: "Can you give an example?", "How did that make you feel?", "Did this change over time?", "Was this related more to academic communication or social communication?", "How did this compare with practising English with classmates, teachers, or friends?", and "Can you explain why that aspect was helpful or unhelpful?"

**Closing script**

Thank you for sharing your experiences. Your responses will be anonymised and used only for research purposes. You may contact the researcher after the interview if you wish to add, clarify, or withdraw any part of your response within the agreed withdrawal period.